\documentclass[aps,pra,showpacs, twocolumn]{revtex4}

\usepackage{graphicx}
\usepackage{amssymb}
\usepackage{amsmath}

\newcommand{\vep}{{\varepsilon}}
\newcommand{\PT}{{$\cal PT$}}

\newcommand{\tw}{\tilde{w}}
\newcommand{\tb}{\tilde{b}}
\newcommand{\tpsi}{\tilde{\psi}}

\newcommand {\Image}{\textrm{Im\ }}
\newcommand {\sech}{\textrm{sech}}

\newcommand{\bp}{{\bf p}}

\begin{document}

\title{Nonlinear modes in the harmonic \PT-symmetric potential}

\author{Dmitry  A. Zezyulin$^{1}$ and Vladimir V. Konotop$^{1,2}$}

\affiliation{
$^1$Centro de F\'isica Te\'orica e Computacional,   Faculdade de Ci\^encias, Universidade de Lisboa, Avenida Professor Gama Pinto 2, Lisboa 1649-003, Portugal
\\
$^2$ Departamento de F\'isica, Faculdade de Ci\^encias,
Universidade de Lisboa, Campo Grande, Ed. C8, Piso 6, Lisboa
1749-016, Portugal
}

\date{\today}

\begin{abstract}
We  study the families of nonlinear modes  described by the
nonlinear Schr\"odinger equation with the \PT-symmetric harmonic
potential $x^2-2i\alpha x$. The found nonlinear  modes display a
number of interesting features. In particular, we have observed
that the  modes, bifurcating  from the different eigenstates of
the underlying linear problem,  can actually belong to the same
family of nonlinear modes. We also show that by proper adjustment of the coefficient $\alpha$
it is possible to enhance stability of small-amplitude and
strongly nonlinear modes comparing to the well-studied case of the
real harmonic potential.
\end{abstract}


\maketitle

\section{Introduction}
\label{intro}

The interest in the stationary modes of the nonlinear
Schr\"odinger equation  with a potential has been raised about two
decades ago  in connection with the applications to the meanfield
dynamics of Bose-Einstein condensates~\cite{firstBEC} and later on in the context of in the context of optical applications~\cite{Kunze} and in particular of propagation of dispersion managed solitons in fibers~\cite{TurMez}. Various
aspect of the nonlinear modes  in a parabolic trapping potential
with homogeneous~\cite{nonlin_homo,AlfZez07,AlZez,PelKevr2010} and
inhomogeneous~\cite{nonlin_inhomo} nonlinearities have been
intensively studied.  A comprehensive analysis of the structure of
the nonlinear modes  and their stability can be found
in~\cite{AlfZez07,AlZez,PelKevr2010}. Further, accounting that
interaction of a particle with a potential in practice is not
absolutely elastic, and energy losses are possible,
in~\cite{(1-i)x^2} there have been addressed  nonlinear modes in a
complex parabolic potential $(1-i)x^2$ supported by a homogeneous
gain. Due to its  dissipative nature, the complex parabolic
potential has properties very different  comparing to its real
counterpart. In particular, for the fixed parameters of the
dissipative  model, the stable nonlinear modes appear as isolated
attractors and do not constitute  continuous families.
Another interesting feature of the complex parabolic potential  is
that in the  limit of the strong defocusing (or repulsive)
nonlinearity, the so-called Thomas-Fermi approximation of the
model is described by the balance between the losses and the gain.
This is not the case of the conservative potential, where the
behavior of the nonlinear  modes in the Thomas-Fermi limit is
determined by the balance between the dispersion (or diffraction)
and the nonlinearity.

In the meantime, recently there has appeared a rapidly increasing
interest~\cite{special}   in linear and nonlinear properties
of the systems with potentials obeying the so-called \PT
~symmetry.  This interest was initiated by the paper~\cite{first},
and more recently by the experimental observation
of \PT~symmetry breaking in optics~\cite{experiment}, as well as
by several theoretical suggestions of realization of \PT-symmetric optical
systems~\cite{suggestions}.

The nonlinear extensions of the \PT-symmetric structures have been
first considered in~\cite{BBJ}.  Later on the nonlinear modes have been studied in the periodic~\cite{periodic},
Gaussian~\cite{ChineseGauss}, and
$\sech^2$-shaped~\cite{ChineseSech} \PT-symmetric potentials, as
well as in the harmonic trap with  rapidly decaying
\PT-symmetric imaginary component~\cite{ChineseBEC}. We also
mention studies of gap solitons in \PT-symmetric optical lattices
combined with real superlattices~\cite{superlattice} and optical
defect modes in \PT-symmetric potentials~\cite{defect}. The modes
and their stability in the systems with  \PT-symmetrically
modulated nonlinearity landscapes have been recently reported
in~\cite{PT-nonlin}.

It turns out, however, that nonlinear modes  in the \PT-symmetric
parabolic trap have not received any attention, so far, while
such a potential, namely $(x-i\alpha)^2$, has been introduced and well
studied in the linear theory~\cite{Kato, Znojil99, BenderJones08}. Meantime,  as
it will be shown below, the  nonlinear modes in the \PT-symmetric
harmonic potential  display  rather unusual properties, which
cannot be observed either in conservative or in
dissipative  potentials of a general kind. The main goal aim of the present work is to
perform a detail study of such modes.

The rest of the paper is organized as follows. In the next section
we introduce the nonlinear model with the \PT-symmetric harmonic
potential and briefly discuss its physical relevance. In
Sec.~\ref{sec:linear} we discuss some properties of the underlying
linear model. Next, in Secs.~\ref{sec:nonlin} and \ref{sec:stab}
and we report the families of nonlinear modes, as well as a detail
investigation of their stability. Sec.~\ref{sec:Concl} concludes
the paper.

\section{The  main model}
\label{sec:model}%
Our main object in the paper is the nonlinear Schr\"{o}dinger equation with a
\PT-symmetric  parabolic potential:
\begin{eqnarray}
\label{CNLS} i q_z =-  q_{xx} + (x^2- 2i\alpha x)q - \sigma|q|^2 q,
\end{eqnarray}
where $\alpha \geq 0$ and $\sigma=1$ and $\sigma=-1$ correspond to
focusing and defocusing nonlinearity (hereafter we use terminology
relevant to optical applications). Physically the dimensionless
Eq.~(\ref{CNLS}) naturally appears as an equation modeling a beam guidance in
a medium whose refractive index $n(x)=n_r(x)+in_i(x)$ has
parabolic modulation of the real part $n_r(x)=x^2$ and linear
modulations of the imaginary part $n_i(x)=2\alpha x$. Even more
generally, any smooth enough symmetric profile of the refractive
index $n_r(x)=n_r(-x)$ and anti-symmetric modulation of its
imaginary part, $n_i(x)=-n_i(-x)$ leads to the model (\ref{CNLS})
if a guided beam is narrow enough allowing  for the use of the
first order terms of the Taylor expansion of the complex index
$n(x)$.

In this paper we are interested in  stationary modes, which  are
searched in the form $q(z, x)=w(x)e^{i\beta z}$, where  $\beta$ is
the propagation constant. We consider localized solution
which obey the zero boundary conditions:
\begin{eqnarray}
\label{boundary}
\lim_{|x|\to\infty}|q(z, x)|=0.
\end{eqnarray}

For the next consideration it is convenient to introduce the
representation  $\beta = b - \alpha^2$, where $b$ is a new
parameter, which allows one to arrive at the following  stationary
equation:
\begin{eqnarray}
\label{nonlin_stat}
w_{xx}- bw  - (x-i\alpha)^2w  +  \sigma|w|^2 w=0.
\end{eqnarray}

Recalling that the existence of the modes implies the balance between the diffraction and the nonlinearity, as well as between gain and losses, we also rewrite Eq.~(\ref{nonlin_stat}) in  the hydrodynamical from
\begin{subequations}
\label{hydro}
\begin{eqnarray}
\label{hydro1}
&&\rho_{xx}- (b-\alpha^2-x^2)\rho  +  \sigma\rho^3- \frac{j^2}{\rho^3}=0,
\\
\label{hydro2}
&& j_x=-2\alpha x\rho^2,
\end{eqnarray}
\end{subequations}
where $\rho(x)=|w(x)|$ is the field modulus, while
$j(x) = \theta_x(x)\rho^2(x)$, with $\theta(x)=\arg w(x)$, is the real-valued current.
From (\ref{hydro}) one readily concludes  that both   $\rho(x)$  and $j(x) = \theta_x(x)\rho^2(x)$
are even functions. The current $j(x)$ has a local maxima at $x=0$, while $\rho(x)$ has  either a local maximum or a local minimum at $x=0$. Moreover, it follows from Eq.~(\ref{boundary}) that $j\to 0$ at $x\to \infty$, and hence taking into account that $j_x(x)$ does not change sign  for $x \ne 0$, we deduce from Eq.~(\ref{hydro1}) that $j(x)$ does not become zero at any finite  $x$, and hence the same is valid to $\rho(x)$ [since otherwise the last term in Eq.~(\ref{hydro1}) would give a singularity]. The absence of zeros of the field contrasts to the known behavior of the nonlinear modes in a real harmonic potential, while is known for the linear \PT-symmetric modes~\cite{Znojil99}, which are briefly outlined in the next section.

\section{Linear modes}
\label{sec:linear}%
Let us recall some relevant properties of the
linear problem~\cite{Kato, Znojil99, BenderJones08}
\begin{eqnarray}
\label{eq:lin}
{\cal L}_n\tw_n=0,
\qquad
{\cal L}_n = \frac{d^2\ }{d x^2}  - \tb_n - (x-i\alpha)^2,
\end{eqnarray}
which can be formally obtained by
setting $\sigma=0$ in Eq.~(\ref{nonlin_stat}).
Hereafter a tilde distinguishes solutions of  the linear
problem.  The set of the eigenvalues of the problem
(\ref{eq:lin}) does not depend on $\alpha$ and consists of an
equidistant sequence  $\tb_n = -(2n+1)$, $n=0,1, \ldots$.
Corresponding eigenfunctions  can be written as $\tw_n(x)= c_n
\tpsi_n(x-i\alpha)$, where $\tpsi_n(x)=H_n(x)e^{-x^2/2}$ is the
$n$th Gauss-Hermite mode, $\int \tpsi_n(x) \tpsi_m^*(x) dx =
\delta_{n,m} \sqrt{\pi}2^nn!$,  $H_n(x)$ is the $n$th Hermite polynomial, and $c_n$ are the positive
coefficients providing the normalization  condition $\int \tw_n(x)
\tw_n^*(x)  dx = 1$ (hereafter we omit the integration limits wherever  the integration is
over whole real axis, and the asterisk denotes complex-conjugation).

Unlike in the conservative case $\alpha=0$, for
$\alpha > 0$ the eigenfunctions $\tw_n(x)$ are not orthogonal.
Using the relation (see e.g.~\cite{MF}): $H_n(x+x_0)=\sum_{k=0}^n{C_n^k} (2x_0)^{n-k}H_k(x)$
where $C_n^k=n!/[k!(n-k)!]$ are the binomial
coefficients), for any $n$ and $m$ one finds
\begin{eqnarray*}
&\int \tw_n(x)\tw^*_m(x) dx
=c_nc_me^{\alpha^2}\sqrt{\pi}\times&\\[2mm]
&\sum_{k=0}^{p}
C_n^kC_m^k 2^k k! (-1)^{n-k} (2 i \alpha)^{n+m-2k}=&\\[2mm]
&c_nc_me^{\alpha^2}\sqrt{\pi}2^{\frac{n+m+g}{2}}i^{3n+m}\alpha^g p!L_{p}^{(g)}(-2\alpha^2),&
\end{eqnarray*}
where $p=\min(n,m)$, $g = |n-m|$, and $L_p^{(g)}(x)$ is the generalized  Laguerre polynomial.
Setting  $n=m$ we obtain the expression for the normalization
coefficients $c_n$:
\begin{equation}
\label{eq:coefs}
c_n = \frac{e^{-\alpha^2/2}}{\sqrt{\sqrt{\pi}2^nn!L_n(-2\alpha^2)}}.
\end{equation}

For $\alpha=0$ the eigenfunctions $\tw_n(x)$  are real-valued (up
to irrelevant phase shift). Moreover, $\tw_n(x)$ is an even (odd)
function if $n$ is even (odd). For $\alpha\ne 0$  the
eigenfunctions are complex-valued  and are neither even nor odd.
Instead, they can be chosen to have even real part and odd
imaginary part.

\section{Bifurcations of nonlinear modes}
\label{sec:nonlin}
 Turning now to the nonlinear problem, we
observe that the eigenvalues $\tb_n$, $n=0,1,\ldots$ are the
bifurcation points where families of nonlinear modes branch off
from the zero solution $w(x)\equiv 0$.  The nonlinear modes
$w_n(x)$  belonging to the $n$th family have the same symmetry as
the corresponding linear eigenfunction $\tw_n(x)$. In the vicinity
of the $n$-th bifurcation point, the nonlinear modes $w_n(x)$
 can be described by means of asymptotic expansions
 \begin{equation}
 \label{eq:expans}
w_n(x)=\vep\tw_n + \mathcal{O}(\vep^3), \quad b_n = \tb_n +
\sigma\vep^2b_n^{(2)} + o(\vep^2),
\end{equation}
 where  $\vep\ll 1$ is a formal
small parameter. Since $\tw_n(x)$ were chosen  normalized, in the
leading order the total energy flow $U=\int |w_n(x)|^2 dx$ (hereafter all the integrals are taken over the whole real axis), is
equal to $\vep$: $U\sim\vep^2$. The  solvability condition for the
$\vep^3$-order equation yields
\begin{equation}
\label{b2}
b_n^{(2)} = \frac{\int \tw_n^3(x)\tw_n^*(x)dx}{\int \tw_n^2(x)dx}.
\end{equation}

Since for $\alpha=0$ the eigenfunctions $\tw_n(x)$  are real
valued,  one has that   $b_n^{(2)}>0$ for any $n$. For $\alpha>0$
the eigenfunctions $\tw_n(x)$ are complex-valued. However, parity
of their real and imaginary parts ensures that $b_n^{(2)}$ is
nevertheless real for any $n$ and $\alpha$. It is straightforward
to obtain explicit expressions for  $b_n^{(2)}$. For the two
lowest families ($n=0$ and $n=1$) one has
\begin{eqnarray}
\label{eq:bn2}
b_0^{(2)} = \frac{e^{\frac{1}{2}\alpha^2}}{\sqrt{2\pi}}, \quad
b_1^{(2)} = \frac{3e^{\frac{1}{2}\alpha^2}}{4\sqrt{2\pi}}\frac{1+2\alpha^2
- \alpha^4}{1+2\alpha^2}.
\end{eqnarray}
It follows from  Eqs.~(\ref{eq:bn2}) that $b_0^{(2)}$ is positive
for all $\alpha$ while $b_1^{(2)}$ is positive  for small $\alpha$, but becomes  negative for $
\alpha > \sqrt{1+\sqrt{2}}$. Regarding the next families, we have found
that for $n=2$  the coefficient $b_2^{(2)}$
 changes sign twice. For $n=3$, however, the coefficient $b_3^{(2)}$ changes sign only once, becoming negative for all
sufficiently large $\alpha$.

From  Eqs.~(\ref{eq:bn2}) we also arrive at another interesting
observation:  the coefficients $b_{0,1}^{(2)}$ grow exponentially
fast with $\alpha$. This, in particular, means that
$\lim_{\alpha\to\infty} \left.\frac{\partial U}{\partial
b}\right|_{b=\tb_{0,1}}=0$. Taking into account that   the
coefficients $b_{0}^{(2)}$ and $b_{1}^{(2)}$  have opposite signs
for $\alpha\gg 1$, one can expect that  for large $\alpha$ the
nonlinear modes bifurcating from  $\tb_0$ and $\tb_1$  merge (or
intersect) at some value of the energy flow  $U$.

\begin{figure}
\centering
        \includegraphics[width=\columnwidth]{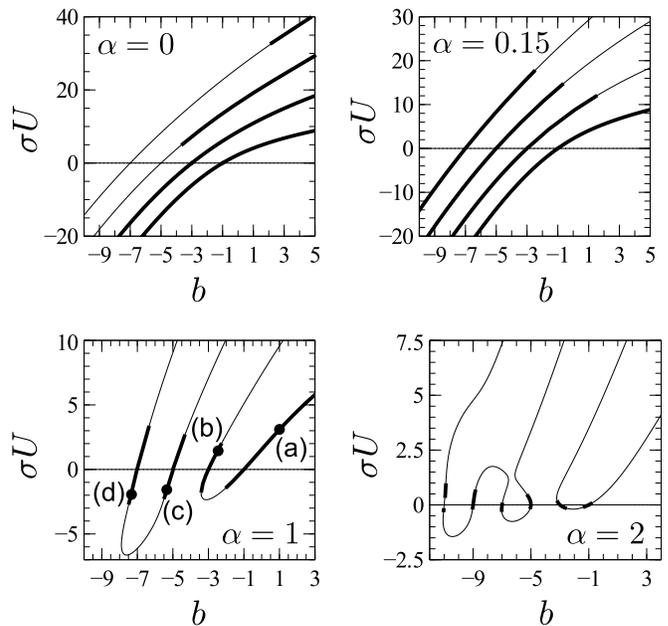}
\caption{The lowest families of nonlinear modes for  different
$\alpha$. The fragments of curves corresponding to stable
nonlinear modes are shown in bold. The  nonlinear modes indicated
with the points (a)--(d) in the panel $\alpha=1$ are explicitly shown in
Fig.~\ref{fig-profs}.} \label{fig-fams}
\end{figure}

The latter situation  seems to be counterintuitive and strongly
contrasting to what is known for the   conservative harmonic
potential, where the modes  bifurcating from different eigenstates
of the linear problem do not merge. In order to check this issue
we performed the direct numerical study  of the families of
nonlinear modes. The characteristic results are summarized in
Fig.~\ref{fig-fams}, where the families of nonlinear modes are shown
on the plane $(b, \sigma U)$ for several different values of
$\alpha$. Respectively, the modes corresponding to the focusing
(defocusing) nonlinearity are situated above (below) the axis
$\sigma U=0$, which is indicated with the dashed line.

\begin{figure}
\centering
        \includegraphics[width=\columnwidth]{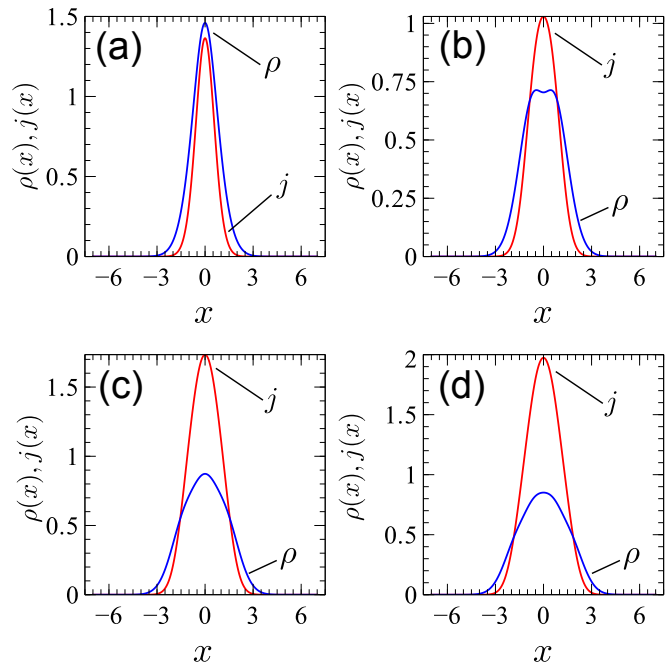}
\caption{(Color online) The modulus $\rho(x)$ and  the current $j(x)$
for stable nonlinear modes corresponding to $\alpha=1$. Panels
(a)--(d) correspond to nonlinear modes indicated by  the points
(a)--(d) in the panel $\alpha=1$ of Fig.~\ref{fig-fams}.}
\label{fig-profs}
\end{figure}

For the sake of comparison,  in the left upper panel of
Fig.~\ref{fig-fams}  we show the  families of nonlinear modes for the
well-studied real  harmonic
oscillator~\cite{nonlin_homo,AlZez,PelKevr2010},  which in our
case  corresponds  to $\alpha=0$.  Increasing $\alpha$ (see the
other panels of Fig.~\ref{fig-fams}), we observe that already at
$\alpha=1$ in the defocusing medium the nonlinear modes
bifurcating from $\tb_0=-1$ and $\tb_1=-3$ (as well as the ones
bifurcating  from $\tb_2=-5$ and $\tb_3=-7$)  indeed appear to be
connected in a single family.For larger  $\alpha$ (e.g. for $\alpha=2$) the structure of
the nonlinear modes  becomes more complicated and the higher
families (the ones bifurcating from $\tb_4=-9$ and $\tb_5=-11$)
also turn to be involved in creation of a singe family snaking
through the linear eigenstates with $n=2,3,4$, and $5$. For
$\alpha=2$ one can see the connection of the modes not only in the
defocusing medium but also in the focusing one.  Since  $\alpha=2>\sqrt{1+\sqrt{2}}$,   Eqs.~(\ref{eq:bn2}) imply that the coefficient $b_1^{(2)}$ is negative
for $\alpha=2$, and thus,  in contrast to the cases $\alpha=0$,
$\alpha=0.15$ and $\alpha=1$, the slope $\partial(\sigma
U)/\partial b$ is negative in the vicinity of the bifurcation from
the point $\tb_1$. In Fig.~\ref{fig-profs} we show the  field modulus
$\rho(x)$ and the superfluid current $j(x)$ for several stable
nonlinear modes corresponding to $\alpha=1$. In accordance with the  discussion in Sec.~\ref{sec:model}, both  $\rho(x)$ and $j(x)$  are even functions, and for all the shown  modes  the current $j(x)$ has a maximum at $x=0$. The field modulus $\rho(x)$ has a maximum at $x=0$  for the nonlinear modes (a), (c) and (d).  For the nonlinear mode (b) the field modulus has a local minimum at $x=0$.

It is interesting to observe, that the described behavior of the modes allows one to suggest  that it is possible to  use continuous deformation to transform one of the modes of the conventional linear harmonic oscillator to another one having different parity. Indeed, to this end it is enough to properly change the strength of the nonconservative potential $\alpha$ and the intensity of the beam $U$. Notice that the stability of the modes, important for any practical realization of such a deformation is discussed in the next section.

\section{Stability of the nonlinear modes}
\label{sec:stab}
\subsection{Analytical results}
Now we turn to analysis of the linear stability of the modes.  Following to
the standard procedure, we use the  substitution  $q(z,x) =
e^{i\beta z}[w(x) +u(x)e^{i\omega z} + v^*(x)e^{-i\omega^*
    z}]$ and arrive at the   eigenvalue problem
\begin{equation}
    \label{BdG}
    {\bf L}\, \bp  = \omega\, \bp ,
\end{equation}
where
\begin{eqnarray*}
        \label{eq:L}
    {\bf L} = \left(
    \!\!
    \begin{array}{cc}%
    L + 2\sigma|w_n|^2 & \sigma w_n^2
    \\
    -\sigma (w_n^2)^* & -L^\dag - 2\sigma|w_n|^2%
    \end{array}
    \!\!\right),\quad
    {\bf p} =\left(
    \!
    \begin{array}{c}%
    u
    \\%
    v
    \end{array}
    \!\right),
\end{eqnarray*}
$L=d^2/dx^2 - b - (x-i\alpha)^2$,  and $L^\dag$ is the Hermitian
adjoint operator. The nonlinear mode $w_n(x)$ is unstable if there
exists an eigenvalue $\omega$ such that  $\Image \omega <0$.

It is straightforward to check the properties of the operator ${\bf L}$ as follows. If $\omega$ is an eigenvalue of ${\bf
L}$ with an eigenvector $(u(x), v(x))^T$, then $-\omega^*$  is also an eigenvalue with  an eigenvector $(v^*(x), u^*(x))^T$. Employing the symmetry of the nonlinear modes
[$w_n(x)=w_n^*(-x)$] one   finds that $\omega^*$ is also an
eigenvalue with an eigenvector $(u^*(-x), v^*(-x))$. Also, $\omega=0$ is  always an
eigenvalue of the operator ${\bf L}$. A corresponding to $\omega=0$ eigenvector
reads $(w_n(x), -w_n^*(x))^T$.

Let us now analyze the spectrum of the operator ${\bf L}$ in the
vicinity of the $n$th bifurcation point. In the linear
limit (i.e. for $\vep=0$) the operator ${\bf L}$ acquires the form
\begin{equation}
    {\bf L} = \tilde{{\bf L}}_n\equiv \left(\begin{array}{cc}%
    {\cal L}_n & 0\\%
    0 & -{\cal L}_n^\dag%
    \end{array}\right),
\end{equation}
where ${\cal L}_n$ is defined in (\ref{eq:lin}). The spectrum of
the operator $\tilde{{\bf L}}_n$ consists of two sequences.
Eigenvalues and eigenvectors of the first sequence read
$\omega^{(I)}_{n,k} = 2(n-k)$, $\bp^{(I)}_{n,k} = (\tw_k(x),
0)^T$, $k=0,1,\ldots$. The second sequence reads
$\omega^{(II)}_{n,k} = -2(n-k)$, $\bp^{(II)}_{n,k} = (0,
\tw_k^*(x))^T$, $k=0,1,\ldots$. First, we notice that the operator
$\tilde{{\bf L}}_n$  has a double zero eigenvalue
$\omega^{(I)}_{n,n} = \omega^{(II)}_{n,n}=0$. Generically, passing
from the linear limit $\vep=0$   to $\vep>0$, a double eigenvalue
splits into two simple eigenvalues. However, in the case at hand,
the splitting of the double zero eigenvalue can not occur. Indeed,
if the zero eigenvalue splits into two simple ones,   they will be
either both real and of opposite signs or complex conjugated.
Either of  these possibilities  means that for $\vep \ne 0$ the
eigenvalue $\omega=0$ is no longer in the spectrum of the operator
${\bf L}$. This, however, contradicts to the established above
properties of the operator ${\bf L}$. Thus, for $\vep \ne 0$ the
operator ${\bf L}$ also has the double zero eigenvalue.

Besides of  the double zero eigenvalue,   the operator
$\tilde{{\bf L}}_n$  has $2n$ double eigenvalues: $\Omega_{n,k} = \omega^{(I)}_{n,k} = \omega^{(II)}_{n,2n-k}$, where $k$ runs from 0 to $2n$ except for $k=n$.  Again,  the double eigenvalue $\Omega_{n,k}$  generically splits
into two simple eigenvalues, which will  be either both real or
complex conjugated. At the same time, the opposite double
eigenvalue $\Omega_{n,2n-k} = -\Omega_{n,k}$ will split in the same manner. Since the
double eigenvalues  $\Omega_{n,k}$ and $\Omega_{n,2n-k}$ behave in the same way,
it is sufficient to analyze only $n$ positive double eigenvalues $\Omega_{n,k}$ which correspond to
 $k=0, \ldots, n-1$.  The double eigenvalue $\Omega_{n,k}$ is
semi-simple; the corresponding eigenvectors  read $\bp^{(I)}_{n,k}
= (\tw_k, 0)^T$ and  $\bp^{(II)}_{n,2n-k} = (0, \tw_{2n-k}^*)^T$.

In order to examine splitting of the double  eigenvalues, we employ Eqs.~(\ref{eq:expans}), which yield the  following asymptotic expansion for the linear stability operator:
${\bf L} =  \tilde{{\bf L}}_n + \sigma \vep^2 {\bf L}_n^{(2)} +
o(\vep^2)$, where
\begin{equation}
    {\bf L}_n^{(2)} = \left(\begin{array}{cc}%
    -\tb_n^{(2)} + 2|\tw_n|^2&  \tw_n^2\\%
     -(\tw^2_n)^*& \tb_n^{(2)} - 2|\tw_n|^2%
    \end{array}\right).
\end{equation}
Following the standard arguments of the perturbation theory for
linear operators~\cite{Kato}, in order to explore the behavior of
a  double eigenvalue $\Omega_{n,k}$ we introduce a $2\times2$   matrix
\begin{eqnarray*}
     {\bf M}_{n,k} = \left(
     \!\!
     \begin{array}{cc}%
    \frac{\langle  {\bf L}_n^{(2)}\bp^{(I)}_{n,k}, {\bp^{(I)*}_{n,k}}\rangle}{\langle\bp^{(I)}_{n,k}, {\bp^{(I)*}_{n,k}}\rangle}&%
    \frac{\langle  {\bf L}_n^{(2)}\bp^{(II)}_{n,2n-k}, {\bp^{(I)*}_{n,k}}\rangle}{\langle\bp^{(I)}_{n,k}, {\bp^{(I)*}_{n,k}}\rangle}\\[6mm]%
    \frac{\langle  {\bf L}_n^{(2)}\bp^{(I)}_{n,k}, {\bp^{(II)*}_{n,2n-k}}\rangle}{\langle\bp^{(II)}_{n,2n-k}, {\bp^{(II)*}_{n,2n-k}}\rangle}&%
    \frac{\langle  {\bf L}_n^{(2)}\bp^{(II)}_{n,2n-k}, {\bp^{(II)*}_{n,2n-k}}\rangle}{\langle\bp^{(II)}_{n,2n-k}, {\bp^{(II)*}_{n,2n-k}}\rangle}
    \end{array}
    \!\!
    \right),
\end{eqnarray*}
where $\langle {\bf a}, {\bf b}\rangle=\int {\bf b}^\dag(x) {\bf a} (x)dx$ for any two column vectors ${\bf a}$ and ${\bf b}$. If both the eigenvalues of the matrix  ${\bf M}_{n,k}$ are real,
then the simple eigenvalues emerging from $\Omega_{n,k}$ are
real, at least for $\vep\geq 0$ sufficiently small. If such a
situation takes place for all $k=0,1,\ldots, n-1$, then one can
state that  the nonlinear modes $w_n(x)$ belonging to the $n$th family are
stable in the linear limit.  On the other hand, if  for some $k$
the matrix ${\bf M}_{n,k}$ has a complex eigenvalue, then the
double eigenvalue $\Omega_{n,k}$ gives rise to a pair of  complex
conjugated eigenvalues. This is sufficient to conclude that  the
nonlinear modes of the $n$th family are unstable in the linear
limit. For $n=0$ no  double eigenvalues $\Omega_{n,k}$ exists. Therefore the lowest family $n=0$ is always stable in the linear limit.

Taking into account symmetry of the eigenfunctions $\tw_n(x)$, one finds that the entries of
the matrix ${\bf M}_{n,k}$ have the form:
\begin{eqnarray*}
\left( {\bf M}_{n,k}\right)_{1,1} = -b_n^{(2)} + 2\frac{\int|\tw_n|^2\tw_k^2
dx}{\int \tw_k^2dx},\\
\left( {\bf M}_{n,k}\right)_{2,2} = b_n^{(2)} -
2\frac{\int|\tw_n|^2\tw_{2n-k}^2 dx}{\int \tw_{2n-k}^2dx},\\
\left( {\bf M}_{n,k}\right)_{2,1} = -\frac{\int\tw_n^2\tw_k^* \tw_{2n-k} dx}{\int
\tw_{2n-k}^2dx},\\
\left(  {\bf M}_{n,k}\right)_{1,2} = \frac{\int\tw_n^2\tw_k \tw_{2n-k}^* dx}{\int
\tw_{k}^2dx}
\end{eqnarray*}
One also observes that all these entries are real.

Using the above expressions, the matrices   ${\bf M}_{n,k}$  as
well as their eigenvalues can be found explicitly.
One observes that for any $n$ and $k$ an expression for
the eigenvalues of the matrix   ${\bf M}_{n,k}$ contains a term
$\sqrt{P_{n,k}(\alpha)}$, where $P_{n,k}(\alpha)$ is a polynomial with real coefficients. Such polynomials are  different for different $n$ and $k$, and are computable explicitly. Their properties (for $n=1,2\ldots 5$) are summarized in Table~\ref{tbl-Pnk}.

The terms $\sqrt{P_{n,k}(\alpha)}$ represent the only possibility for the eigenvalues eventually to have a nonzero imaginary part. Respectively, splitting of the double  eigenvalue  $\Omega_{n,k}$  for $\alpha=0$ is determined
by the sign of $P_{n,k}(0)$, while the behavior of  $\Omega_{n,k}$  in the limit $\alpha \gg 1$ is determined by the sign
of the leading coefficient of the polynomial $P_{n,k}(\alpha)$.

\begin{table}
    \begin{tabular}{c|c|c|c|c|c@{\hspace{2mm}}c@{\hspace{2mm}}c@{\hspace{2mm}}c@{\hspace{2mm}}c@{\hspace{2mm}}c}
    $n$     &  $k$      &  $D$       &    $S$     & $s$   & \multicolumn{6}{|c}{positive roots}\\\hline
    $1$   &  $0$    &  12      &   $+$            &$+$    &\multicolumn{6}{|c}{no positive roots}\\\hline%
    $2$   &  $0$    &  24      &   $+$            &$-$              &0.05&2.47&2.54&3.21&3.60& \\
        &  $1$      &  20      &   $+$            &$+$    &\multicolumn{6}{|c}{no positive roots}\\\hline
    $3$   &  $0$    &  36      &   $+$            &$-$              &0.12&&&&&\\
          &  $1$    &  32      &   $+$            &$-$              &0.05&1.68& 1.94&3.18&4.17&  \\
          &  $2$    &  28      &   $+$            &$+$    &\multicolumn{6}{|c}{no positive roots}\\\hline
    $4$   &  $0$    &  48      &   $+$            &$+$              &0.08&0.14&3.35&3.40&4.77&4.82\\
          &  $1$    &  44      &   $+$            &$-$              &0.11&&&&&\\
          &  $2$    &  40      &   $+$            &$-$              &0.05&3.64&4.66&&&\\
          &  $3$    &  36      &   $+$            &$+$    &\multicolumn{6}{|c}{no positive roots}\\\hline
    $5$   &  $0$    &  60      &   $+$            &$+$              &0.12&0.14&&&&\\
          &  $1$    &  56      &   $+$            &$-$              &0.12&1.74&2.20&5.14&5.24&\\%
          &  $2$    &  52      &   $+$            &$-$              &0.10&2.41&2.58&&&\\
          &  $3$    &  48      &   $+$            &$-$              &0.05&2.84&2.92&4.10&5.10&\\
          &  $4$    &  44      &   $+$            &$+$    &\multicolumn{6}{|c}{no positive roots}
    \end{tabular}
    \caption{Properties of the polynomials $P_{n,k}(\alpha)$.
    Here $D_{n,k}$ is the degree of a polynomial, $S_{n,k}$ is the sign \
    of the leading coefficient, $s_{n,k}$ is the sign of the constant
    term $P_{n,k}(0)$. Approximate values of all the positive roots are also
    reported.  }
    \label{tbl-Pnk}
    \end{table}

Turning now to the Table~\ref{tbl-Pnk}, the following comments can
be given: (i) the degree of the  polynomial $P_{n,k}(\alpha)$, denoted by $D_{n,k}$,
obeys the relation $D_{n,k}=12n-4k$; (ii) more importantly, the leading
coefficients of all the considered polynomials are positive. It
means that for any $n$ there exists a critical value
$\alpha_n^{cr}$  such that for all $\alpha>\alpha_n^{cr}$ the
$n$th family is stable in the linear limit, even if this family is
unstable in the  case of the real harmonic potential (i.e. for $\alpha=0$). For $n=0$ and $n=1$ the critical values
are zero: $\alpha_0^{cr}=\alpha_1^{cr}=0$. Respectively, these
families are stable in the linear limit both for the case of the
real harmonic oscillator ($\alpha=0$), and in \PT-symmetric case
for any $\alpha$. For the next families, $n=2$ and $n=3$,   the Table~\ref{tbl-Pnk}   yields $\alpha_2^{(cr)}\approx 3.60$,  and $\alpha_3^{(cr)}\approx 4.17$. This
means that being unstable in the linear limit for $\alpha=0$, the latter families
become stable in the linear limit for $\alpha$ sufficiently large. The same  situation
takes place to the  families $n=4$ and $n=5$. Moreover we conjecture that it also
persists for all higher families.

\subsection{Numerical results}
Passing to the numerical study of the stability (see
Fig.~\ref{fig-fams}), we first recall some results  known for the
real harmonic potential, which in our model corresponds to
$\alpha=0$. The  nonlinear  modes that belong to two lowest
families ($n=0$ and $n=1$) are always stable. The families $n=2$
and $n=3$ are unstable in the linear limit and for small and
moderate values of $U$. However both the latter families
become stable if the nonlinearity is  sufficiently strong (for the
stability analysis of the modes in strongly nonlinear defocusing
medium see~\cite{PelKevr2010}).  In the defocusing medium, the
value of $U$, which have to be exceeded    for the families  $n=2$
and $n=3$ to become stable, is large and does not belong to the
scope of the panel $\alpha=0$ of Fig.~\ref{fig-fams}.

In the next panel of Fig.~\ref{fig-fams} we consider the case   $\alpha=0.15$. For this  value of  $\alpha$  it follows from Table~\ref{tbl-Pnk}  that for any $n=1,2,\ldots, 5$ the $n$th family is stable in the linear limit.
Turning to stability of the nonlinear modes of arbitrary amplitude, we observe that the lowest family $n=0$ is stable in the whole the considered  region of parameters. The same situation takes place for the family $n=1$ but in defocusing medium only. For $\sigma=1$ this family loses stability at  sufficiently strong nonlinearity. The most interesting results, however, are  obtained for the families $n=2$ and $n=3$. In contrast to their
counterparts for the real harmonic oscillator, these  families are
{stable in the linear limit}. Moreover,  these  families remain stable  at least for small and moderate values of  $U$.  In the defocusing medium,  the   families $n=2$ and $n=3$  appeared to be stable in the whole the explored region. In the focusing medium, we have found the critical values of nonlinearity after which  onset of instability occurs. It is interesting, that in the certain sense the situations for the real oscillator and  for the \PT-symmetric one are opposite:  for $\alpha=0$ the families $n=2$ and $n=3$ are unstable in the linear limit but become stable  in focusing medium for  $U$ sufficiently  large. {\it Vice versa}, for $\alpha=0.15$, those families  are stable in the linear limit but lose stability  in focusing medium  for large  $U$. At this stage we emphasize  that only finite range of $b$ and $U$ has been considered in our numerics, and in principle, the families of nonlinear modes may change stability for stronger values of nonlinearity,  which have not been considered here.

Next, we have considered \PT-symmetric harmonic potentials with stronger imaginary component, $\alpha=1$ and $\alpha=2$. One can deduce from Table~\ref{tbl-Pnk} that for $\alpha=1$
the families $n=1,2,\ldots 5$ are stable in the linear limit,  while for $\alpha=2$ the families $n=1, \ldots  4$ are stable in the linear limit  and the family  $n=5$ is unstable. However, both for $\alpha=1$ and $\alpha=2$  all the considered families lose stability for relatively small values of $U$.  One observes that  the larger    $\alpha$, the  smaller nonlinearity strength is sufficient for  the destabilization to occur.

\section{Conclusion}
\label{sec:Concl}

To conclude we have performed the analysis of the structure and the stability of the lowest families of nonlinear modes in the nonlinear Schr\"odinger equation with the parabolic \PT-symmetric potential. We have found a number of striking features, not observable for the cases of conservative and dissipative parabolic potentials. Among these features we emphasize  transformation of  the families bifurcating from the different eigenstates of the underlying linear problem  to the single family; enhancement of the stability in the linear limit comparing to the standard case of the real harmonic oscillator; and possibilities of proper choices of the strength of the nonconservative part    $\alpha$  making  unstable for $\alpha=0$ nonlinear modes to become stable in the \PT-symmetric case.

\acknowledgments

DAZ was supported by Funda\c{c}\~{a}o para a Ci\^{e}ncia e a Tecnologia (FCT) under the grant  No. SFRH/BPD/64835/2009. VVK was supported by the FCT under the grant No. PTDC/FIS/112624/2009. The authors acknowledge support by the FCT through the grant PEst-OE/FIS/UI0618/2011.

\end{document}